
%
%
%
%
\font\tenbf=cmbx10

\font\eightrm=cmr8
\font\eightit=cmti8

\def\sectiontitle#1\par{\vskip0pt plus.1\vsize\penalty-250
\vskip0pt plus-.1\vsize\bigskip\vskip\parskip
\message{#1}\leftline{\tenbf#1}\nobreak\vglue 5pt}

\def\eno{\eqalignno}
\def\ld{\lambda}

\def\al{\alpha}

\def\im{\hbox{\rm Im~}}
\def\cot{\hbox{\rm cot~}}

\magnification=\magstep1
\parindent=15pt
\nopagenumbers
\baselineskip=10pt
\line{
hep-th/9403107
\hfil}
\vglue 20pt
\line
{\hfil\it Dedicated to the memory of Ansgar Schnizer}
\vglue 2pc
\baselineskip=13pt
\headline{\ifnum\pageno=1\hfil\else%
{\ifodd\pageno\rightheadline \else \leftheadline\fi}\fi}
\def\rightheadline{\hfil\eightit
Completeness of Bethe's states for generalized $XXZ$ model
\quad\eightrm\folio}
\def\leftheadline{\eightrm\folio\quad
\eightit
Anatol N. Kirillov, Nadejda A. Liskova
\hfil}
\voffset=2\baselineskip
\centerline{\tenbf
COMPLETENESS\hskip 0.1cm OF\hskip 0.1cm BETHE'S \hskip 0.1cm STATES
}
\centerline{\tenbf
FOR \hskip 0.1cm
GENERALIZED \hskip 0.1cm $XXZ$ \hskip 0.1cm MODEL \hskip 0.1cm  \hskip 0.1cm
 }
\vglue 24pt
\centerline{\eightrm
ANATOL N. KIRILLOV
}
\baselineskip=12pt
\centerline{\eightit
Steklov Mathematical Institute,
}
\baselineskip=10pt
\centerline{\eightit
Fontanka 27, St.Petersburg, 191011, Russia
}
\vglue 10pt\centerline{\eightrm
NADEJDA A. LISKOVA
}
\baselineskip=12pt
\centerline{\eightit
St.Petersburg Institute of Aviation Instruments,
}
\baselineskip=10pt
\centerline{\eightit
Gertzena 67, St.Petersburg, 190000, Russia
}
\vglue 15pt
\centerline{\eightrm ABSTRACT}
{\rightskip=1.5pc
\leftskip=1.5pc
\eightrm\parindent=1pc
We study the Bethe ansatz equations for a generalized $XXZ$ model on a
one-dimensional lattice.\break Assuming the string conjecture we propose an
integer version
for vacancy numbers and prove a combinatorial completeness of Bethe's states
for a generalized $XXZ$ model. We find an exact form for inverse matrix
related with vacancy numbers and compute its determinant. This inverse
matrix has a tridiagonal form, generalizing the Cartan matrix of type $A$.}
\vglue12pt
\baselineskip=13pt
\overfullrule=2pt
\def\qed{\hfill$\vrule height 2.5mm width 2.5mm depth 0mm$}
\vskip 0.5cm

{\bf \S 1. Introduction.}
\bigbreak

An integrable generalization of spin-${1\over 2}$ Heisenberg $XXZ$ model to
arbitrary spins was given e.g. in [KR2]. As a matter of fact, a spectrum
of the generalized $XXZ$ model is described by the solutions
$\{\lambda_i\}$ to the following system of equations $(1\le j\le l)$
$$\prod^N_{a=1}{\sinh{\theta\over 2}(\lambda_j+2is_a)\over\sinh{\theta\over
2}(\lambda_j-2is_a)}=\prod^l_{\matrix{k=1\cr k\ne j}}{\sinh{\theta\over
2}(\lambda_j-\lambda_k+2i)\over\sinh{\theta\over
2}(\lambda_j-\lambda_k-2i)}.\eqno (1.1)
$$
Here $\theta$ is an anisotropy parameter, $s_a,~~1\le a\le N$, are the spins
of atoms in the magnetic chain and $l$ is the number of magnons over  the
ferromagnetic vacuum.

The main goal of our paper is to present a computation the number of solutions
of system (1.1) based on the so-called string conjecture (see e.g. [TS],
[KR1]).
In spite of the well-known fact that solutions of (1.1) do not have in
general a ``string nature'' (see e.g. [EKK]), we prove that the string
conjecture gives a
correct answer for the number of solutions to the system of equations (1.1).
Note that a combinatorial completeness of Bethe's states for generalized $XXX$
Heisenberg model was proved in [K1] and appear to be a starting point
for numerous applications to combinatorics of Young tableaux and
representation theory of symmetric and general linear groups (see e.g.
[K2]).

\bigbreak

{\bf \S 2. Analysis of the Bethe equations.}
\medbreak

Let us consider the $XXZ$ model of spins $s_1,\ldots ,s_k$ interacting on
one-dimensional lattice with the each spin $s_i$ repeateded $N_i$ times. In the
standard $XXZ$ model all spins $s_i$ are equal to ${1\over 2}$. Let $\Delta$
be the anisotropic parameter 
(see e.g. [TS], [KR2]). We assume that $0<\theta <1$. Let us pick out a real
number
$\theta$ such that $\cos\theta =\Delta,~~0<\theta <{\pi\over 2}$, and denote
$$p_0={\pi\over\theta }>2.
$$
Each spin $s$ has a ``parity'' $v_{2s}$ which is equal to plus or minus one.

Bethe vectors $\psi(x_1,\ldots ,x_l)$ for $XXZ$ model are parametrized by $l$
complex numbers $x_j({\rm mod~}2p_0i)$ ($l\le 2s_1N_1+\ldots +2s_kN_k$), which
satisfy the following system of transcendental equations (Bethe's equations)
$$\eno{\prod^k_{m=1}(-1)^{N_mv_{2s_m}}&\left({{\sinh}{\theta\over
2}(x_{\alpha}+\eta_m-i(2s_m+{1\over 2}(1-v_{2s_m})
p_0))\over{\sinh}{\theta\over 2}(x_{\alpha}+\eta_m+i(2s_m+{1\over 2}
(1-v_{2s_m})
p_0))}\right)^{N_m}=\cr
&=-\prod^l_{j=1}{{\sinh}{\theta\over 2}(x_{\alpha}-x_j-2i)
\over{\sinh}{\theta\over 2}(x_{\al }-x_j+2i)},&(2.1)\cr
&{\rm where}~~\al =1,\ldots ,l,~~{\rm and}~~\{\eta_m\} ~~{\rm are \  some \
fixed \  real \  numbers},}
$$
and nondegeneracy conditions: the norm of Bethe's vectors $\psi$ is not equal
to zero.

Solutions to the system (2.1) are considered modulo $2p_0i{\bf Z}$, because
${\sinh}({\theta\over 2}x)$ is a periodic function with the period
$2p_0i$.
Asymptotically for $N_m\to\infty,~~1\le m\le k$ and finite $l$ the solutions to
the
system (2.1) create the strings. The strings are characterized by common
real abscissa which is called the string center, the length $n$ and parity
$v_n$. Centers of the
even strings are located on the line ${\rm Im~}x=0$ (and $v_n=+1$), those of
odd strings are located on the line
${\rm Im~}x=p_0$ (and $v_n=-1$). A string of length $n$ and parity $v_n$
consists
of $n$ complex numbers $x^n_{\beta ,j}$  of the following form
$$x^n_{\beta ,j}=x^n_{\beta }+i(n+1-2j+{1-v_n\over 2}p_0)+O(\exp (-\delta N))
({\rm mod~}2p_0i),\eqno (2.2)
$$
where $\delta >0,~~j=1,\ldots ,n,~~x^n_{\beta }\in{\bf R}$.

A distribution of numbers $\{ x_j\}$ on strings is called configuration. Each
configuration can be parametrized by the filling numbers $\{\ld_n\}$,
where $\ld_n$ is equal to the number of strings with length $n$ and parity
$v_n$.
Each real solution of the system (2.1), $({\rm modulo~}2p_0i)$, corresponds to
an
even
string of length $1$. Configuration parameters $\{\ld_n\},~~n\ge 1$ satisfy
the following conditions: $\ld_n\ge 0,~~{\displaystyle\sum_{n\ge 1}n\ld_n=l}$.
The system (2.1) can be transformed into that for real numbers $x^n_{\beta }$
for each fixed
configuration. To get such system, let us calculate the scattering phase
$\theta_{n,m}(x)$ of the string length $n$ on that of length $m$. By definition
$$\exp (-2\pi i\theta_{n,m}(x))=\prod^n_{j=1}\prod^m_{k=1}{\sinh{\theta\over 2}
(x^n_{\al ,j}-x^m_{\beta ,k}-2i)\over\sinh{\theta\over 2}(x^n_{\al ,j}-
x^m_{\beta ,k}+2i)},~~x:=x^n_{\al }-x^m_{\beta }.\eqno (2.3)
$$

{}From the formulae
$$\im\log\left({\sinh (\ld+ai)\over\sinh (\mu+bi)}\right)=\arctan (\tanh
\mu\cdot
\cot b)-
\arctan (\tanh\ld\cdot\cot a),\eqno (2.4)
$$
where $a,b,\ld ,\mu\in{\bf R}$, it follows
$$\eno{
-\pi \theta_{n,m}(x)&=\sum^n_{j=1}\sum^m_{k=1}\arctan\left(\tanh{\theta x\over
2}
\cot{\theta\over 2}(n-m-2j+2k+{1\over 2}(v_m-v_n)p_0)\right)=\cr
&=\arctan (\tanh{\theta x\over 2}\cdot\cot{\theta\over 2}(m+n+{1\over 2}
(1-v_nv_m)p_0))+\cr
&+\arctan (\tanh{\theta x\over 2}\cdot\cot{\theta\over 2}(|n-m|+{1\over 2}
(1-v_nv_m)p_0))+\cr
&+2\sum_{s=1}^{\min (n,m)-1}\arctan (\tanh{\theta x\over
2}\cdot\cot{\theta\over 2}
(|n-m|+2s+{1\over 2}(1-v_nv_m)p_0)).}
$$

Now let us consider the limit of $\theta_{n,m}(x)$ when $x\to\infty$. Note
that for $x\to\infty$, we have
$\tanh{\theta x\over 2}\to 1$, and $\arctan (\cot z)=-\pi(
({z\over\pi}))$, if ${z\over\pi}\not\in{\bf Z}$, where $((z))$ is the Dedekind
function:
$$((z))=\cases{0,&if $z\in{\bf Z}$,\cr
\{ z\} -{1\over 2},&if $z\not\in{\bf Z}$,\cr}
$$
and $\{ z\} =z-[z]$ is the fractional part of $z$. Then
$$\eno{\theta_{n,m}(\infty )&=\left(\left({n+m\over 2p_0}+{1-v_nv_m\over 4}
\right)\right)+\left(\left({|n-m|\over 2p_0}+{1-v_nv_m\over 4}\right)\right)+
\cr
&+2\sum_{l=1}^{
\min (n,m)-1}\left(\left({|n-m|+2l\over 2p_0}+{1-v_nv_m\over 4}\right)\right).}
$$
Let us define
$$\Phi_{n,m}(\ld )=-{1\over 2\pi i}\sum^n_{j=1}\log{\sinh{\theta\over 2}
(x^n_{\al ,j}+\eta -mi-{1\over 2}(1-v_m)p_0i)\over \sinh{\theta\over 2}
(x^n_{\al ,j}+\eta +mi+{1\over 2}(1-v_m)p_0i)},
$$
$$\ld :=x^n_{\al}+\eta,~~\eta\in{\bf R},$$
then
$$\eno{\Phi_{n,m}(\ld )&=-{1\over 2\pi}\sum^n_{j=1}2\cdot\arctan (\tanh{\theta
x\over 2}\cot{\theta\over 2}(n-m+1-2j+{1\over 2}(v_m-v_n)p_0))=\cr
&={1\over\pi}\sum_{l=1}^{\min (n,m)}\arctan (\tanh{\theta x\over
2}\cot{\theta\over 2}(|n-m|+2l-1+{1\over 2}(1-v_nv_m)p_0))}
$$
and, consequently,
$$\Phi_{n,m}(\infty )=\sum_{l=1}^{\min (n,m)}\left(\left({|n-m|+2l-1\over 2p_0}
+{1-v_nv_m\over 4}\right)\right). \eqno (2.5)
$$

Now let us continue the investigation of system (2.1). Multiplying the
equations
of the system (2.1) along the string $x^n_{\al ,j}$ and logarithming the
result, one
can obtain the following system on real numbers $x^n_{\al},~~\al =1,\ldots ,
\ld_n$:
$$\sum_m\Phi_{n,2s_m}(x^n_{\al }+\eta_m)N_m=Q^n_{\al}+\sum_{(\beta ,m)\ne
(\al ,n)}\theta_{n,m}(x^n_{\al}-x^m_{\beta}),~~\al =1,\ldots ,\ld_n. \eqno
(2.6)
$$
Integer or half-integer numbers $Q^n_{\al},~~1\le\al\le\ld_n$, are called
quantum numbers. They parametrize --- according to the string conjecture [TS],
[FT], [K1] ---the solutions to the system (2.1). Admissible values of
quantum numbers $Q^n_{\al}$ are located in symmetric interval $[-Q^n_{\max},~
Q^n_{\max}]$ and appear to be an integer or half-integer in accordance with
that of $Q^n_{\max}$.
\bigbreak

{\bf \S 3. Calculation of vacancy numbers.}
\medbreak

Following [TS], we will assume that there are two types of length $1$ string,
namely even and odd types. If the length $n$ of a string is greater then $1$,
then $n$ and parity $v_n$ satisfy the following conditions
$$\eno{&v_n\cdot\sin (n-1)\theta >0,&(3.1)\cr
&v_n\cdot\sin (j\theta )\sin(n-j)\theta >0,~~j=1,\ldots ,n-1.&(3.2)}
$$
Condition (3.1) may be rewritten equivalently as
$$v_n=\exp (\pi i\left[{n-1\over p_0}\right]),
$$
and (3.2) as
$$\left[{j\over p_0}\right]+\left[{n-j\over p_0}\right]=\left[{n-1\over p_0}
\right],~~j=1,\ldots ,n-1.\eqno (3.3)
$$
The set of integer numbers $n$ satisfying (3.3) for fixed $p_0\in{\bf R}$
may be described by the following construction (see e.g. [TS], [KR1], [KR2]).

Let us define a sequence of real numbers $p_i$ and sequences of integer numbers
$\nu_i,~m_i,~y_i$:
$$\eno{
&p_0={\pi\over\theta},~p_1=1,~\nu_i=\left[{p_i\over p_{i+1}}\right],~p_{i+1}=
p_{i-1}-\nu_{i-1}p_i,~~i=1,2,\ldots &(3.4)\cr
&y_{-1}=0,~y_0=1,~y_1=\nu_0,~y_{i+1}=y_{i-1}+\nu_iy_i,~~i=0,1,2,\ldots
&(3.5)\cr
&m_0=0,~m_1=\nu_0,~m_{i+1}=m_i+\nu_i,~~i=0,1,2,\ldots &(3.6)}
$$

It is clear that integer numbers $\nu_i$ define the decomposition of $p_0$ into
a continuous fraction
$$p_0=[\nu_0,\nu_1,\nu_2\ldots ].
$$
Let us define a piecewise linear function $n_t,~~t\ge 1$
$$n_t=y_{i-1}+(t-m_i)y_i,~~{\rm if}~~m_i\le t<m_{i+1}.
$$
Then for any integer $n>1$ there exist the unique rational number $t$ such that
$n=n_t$.
\medbreak

{\bf Lemma 3.1} [KR1]. {\it Integer number $n>1$ satisfies (3.3) if and only
if there exists an integer number $t$ such that $n=n_t$. }

We have two types of length 1 strings :
$$\eno{
&x_{\al}^1~~{\rm with \ parity}~~v_1=+1,\cr
&x_{\al}^{m_1}~~{\rm with \ parity}~~v_{m_1}=-1.}
$$
All others strings have a length $n=n_j$, for some integer $j$, and parity
$$v_j=v_{n_j}=\exp \left(\pi i\left[\displaystyle{{n_j-1\over p_0}}\right]
\right).
$$

Let us assume that all spins $s_i$ have the following form
$$2s_i=n_{\chi_i}-1,~~\chi_i\in{\bf Z}_+.\eqno (3.7)
$$

{}From the assumptions (3.1),(3.3) and (3.7) about spins, length and parity it
follows
a simple expression for the sums $\theta_{n,m}(\infty )$ and
$\Phi_{n,2s}(\infty )$.

In our paper we consider a special case of rational $p_0$. The case of
irrational
$p_0$ may be obtained as a limit. So, we assume that $p_0={\displaystyle{u\over
v}}\in{\bf Q},~~p_0=[\nu_0,\ldots ,\nu_{\al}],~~\nu_0\ge 2$, \ \
$\nu_{\al}\ge 2$. Furthermore we assume that all strings have a length not
greater than $u$ (see [TS]). Therefore  for numbers $p_i,\nu_i,y_i,m_i$, (see
(3.4)--(3.6)) it is enough to keep only the indices $i\le\al  +1$. We have also
$$p_{\al +1}={\displaystyle{1\over y_{\al}}},~~ p_0={\displaystyle{y_{\al +1}
\over y_{\al}}},~~{\rm and~ g.c.d. }~(y_{\al},~~y_{\al +1})=1.
$$

Now we will state the results of calculations for the sums $\theta_{n,m}(\infty
)$
and $\Phi_{n,m}(\infty )$. Let us introduce
$$\eno{
&q_j=(-1)^i(p_i-(j-m_i)p_{i+1}),~~{\rm if}~~m_i\le j<m_{i+1},\cr
&r(j)=i,~~{\rm if}~~m_i\le j<m_{i+1},\cr
&b_{jk}={(-1)^{i-1}\over p_0}(q_kn_j-q_jn_k),~~{\rm if}~~n_j<n_k,& (3.8)\cr
&b_{j,m_{i+1}}=1, ~~m_i<j<m_{i+1}, \cr
&\theta_{j,k}=\theta_{n_j,n_k}(\infty ),~~\Phi_{j,2s}=\Phi_{n_j,2s}(\infty ).}
$$
\medbreak

{\bf Theorem 3.2.} (Calculation of the sums $\theta_{j,k},~~\Phi_{j,2s}$)
[KR1].

{\it 1. If  $k>j$, and $(j,k)\ne (m_{\al +1}-1,~m_{\al +1})$ then
$$\theta_{jk}=-n_j{q_k\over p_0}.
$$

2. If $j=m_{\al +1} -1,~~k=m_{\al +1}$, then
$$\theta_{jk}=-n_j{q_k\over p_0}+{(-1)^{\al +1}\over 2}=(-1)^{\al}~{p_0-2\over
2p_0}.
$$

3. If $1\le j\le m_{\al +1}$, then
$$\theta_{jj}=-n_j{q_j\over p_0}+{(-1)^{r(j)}\over 2}.
$$

4. (Symmetry). For all \ $1\le j,~~k\le m_{\al +1}$:
$$\theta_{jk}=\theta_{kj}
$$

5. If \ $2s=n_{\chi}-1$, then}
$$\Phi_{k,2s}=\cases{\displaystyle{1\over 2p_0}(q_k-q_kn_{\chi}),&if
$n_k>2s$,\cr \cr
\displaystyle{1\over 2p_0}(q_k-q_{\chi}n_k)+{(-1)^{r(k)-1}\over 2},
&if $n_k\le 2s$.\cr}
$$

Note that if $p_0=\nu_0$ is an integer then
$$2\Phi_{k,2s}=\cases{\displaystyle{2sk\over p_0}-\min (k,2s),& if
$1\le k,~2s+1<\nu_0$,\cr \cr~~0,& if $1\le k\le\nu_0,~2s+1>\nu_0$.}
$$

Now we are going to calculate the vacancy numbers. By definition the vacancy
numbers are equal to
$$\eno{
&P_{n_j}(\ld )=2Q^{n _j}_{\max}-\ld_j+1,~~{\rm where}\cr
&Q^{n_j}_{\max}=(-1)^{i-1}\left(Q^{n_j}_{\infty}-\theta_{jj}-{n_j\over
2}\left\{
{\sum 2s_mN_m-2l\over p_0}\right\}\right)-{1\over 2},~~m_i\le j<m_{i+1},}
$$
and \ $\{ x\}$ is the fractional part of real number $x$.

Here we put
$$Q_{\infty}^{n_j}=\sum_k\Phi_{n_j,2s_k}N_k-\sum_k\theta_{n_jn_k}
\lambda_{k}+\theta_{n_jn_j}.
$$

Let us say few words  about our definition of vacancy numbers $P_{n_j}$.
In contrast with the $XXX$ model situation, it happens that the vector
$x=(\infty ,\ldots ,\infty )$ for the $XXZ$ case does not appear to be a formal
solution to the Bethe equations (2.1). Another difficulty appears in finding a
correct boundary for quantum numbers $Q^n_{\al}$ (see (2.6)). A natural
boundary is $Q_{\infty}^{n_j}$ but this number does not appear to be an
integer or half-integer one in general. Our choice is based on attempt to have
a combinatorial completeness of Bethe's states and some analytical
considerations.
In the sequel we will use the notations $P_j(\ld ), Q^j_{\infty
},Q^j_{\max},\ldots$
instead of $P_{n_j}(\ld ),Q^{n_j}_{\infty}, Q^{n_j}_{\max},\ldots$.

After a tedious calculations one can find
$$\eno{
&P_j(\ld)=a_j+2\sum_{k>j}b_{jk}\ld_k,~~j\ne m_{\al +1}-1,~m_{\al +1},\cr
&P_{m_{\al +1}-1}(\ld )=a_{m_{\al +1}-1}+\ld_{m_{\al +1}},
& (3.9)\cr
&P_{m_{\al +1}}(\ld )=a_{m_{\al +1}}+\ld_{m_{\al +1}-1},}
$$
where
$$a_j=(-1)^{i-1}\left(\sum_m2\Phi_{j,2s_m}\cdot N_m+{2lq_j\over p_0}-n_j
\left\{{\sum 2s_mN_m-2l\over p_0}\right\}\right),
$$
and $b_{jk}$ for $n_j<n_k$ are defined in (3.8).

{}From the string conjecture (see [TS], [KR2]) it  follows that the number of
Bethe's
vectors with configuration $\{\ld_k\}$ is equal to
$$Z(N,s|\{\ld_k\})=\prod_j\pmatrix{P_j(\ld )+\ld_j\cr \ld_j}.
$$
The number of Bethe's vectors with fixed $l$ is equal to
$$Z(N,s|l)=\sum_{\{\ld_k\}}Z(N,s|\{\ld_k\}),\eqno (3.10)
$$
where summation is taken over all configurations $\{\ld_k\}$, such that
$\ld_k\ge 0$, and
$$\sum_{k=1}^{m_{\al +1}}n_k\ld_k=l.\eqno (3.11)
$$
So, the total number of Bethe's vectors is equal to
$$\eno{
&Z=Z(N,s)=\sum_lZ(N,s|l),~~{\rm where~~we~~assume~~that}& (3.12)\cr
&Z(N,s|l):=Z(N,s|\sum 2s_mN_m-l),~~{\rm for}~~l\ge\sum s_mN_m.}
$$
The conjecture about combinatorial completeness of Bethe's states for $XXZ$
model means that
$$Z=\prod_m(2s_m+1)^{N_m}.
$$
\bigbreak

{\bf \S 4. The main combinatorial identity.}
\medbreak

Let $a_0=0,~a_1,a_2,\ldots ,a_{m_{\al +1}}$ be a sequence of real numbers.
Then we shall define inductively a sequence $b_2,\ldots ,b_{m_1-1},
b_{m_1+1},\ldots ,b_{m_{\al +1}},b_{m_{\al +2}}$ by the following rules
$$\eno{
&b_k=2a_{k-1}-a_{k-2}-a_k,~~{\rm if}~~k\ne m_i,~~k\ge 2,\cr
&b_{m_i+1}=2a_{m_i-1}-a_{m_i-2}-a_{m_i+1},~~{\rm if}~~1\le i\le\al ,\cr
&b_{m_{\al +2}}=a_{m_{\al +1}-1}-a_{m_{\al +1}-2}+a_{m_{\al +1}}.}
$$
Then one can check that the converse formulae are
$$a_j=(-1)^{r(j)}\left({n_j\over p_0}q_{m_{\al
+1}}(a_{m-1}-a_m)-2\sum_k\Phi_{jk}\cdot b_k\right),
$$
where $\Phi_{jk}$ were defined in (2.5).

For given configuration $\{\ld_n\}=\ld$ let us define the vacancy numbers
$$\eno{
&P_j(\ld )=a_j+2\sum_{k>j}b_{jk}\ld_k,~~j\ne m_{\al +1}-1,\cr
&P_{m_{\al +1}-1}(\ld)=a_{m_{\al +1}-1}+\ld_{m_{\al +1}},\cr
&P_{m_{\al +1}}(\ld )=a_{m_{\al +1}}+\ld_{m_{\al +1}-1}.}
$$
Let us put
$$Z(\{a_k\}|l)=\sum_{\{\ld_k\}}\prod_{k=1}^{m_{\al +1}}\pmatrix{P_k(\ld )+\ld_k
\cr\ld_k},
$$
where summation is taken over all configurations $\{\ld_k\}$ such that
$$\sum^m_{k=1}n_k\ld_k=l.
$$
Remind that a binomial coefficient $\pmatrix{\al\cr\nu}$ for real $\alpha$
and integer positive $\nu$ is defined as
$$\pmatrix{\al\cr\nu}={\al (\al -1)\ldots (\al -\nu +1)\over \nu !}.
$$

{\bf Theorem 4.1.} (The main combinatorial identity) {\it We have
$$Z(\{ a_k\}|l)={\rm Res}_{u=0}f(u)u^{-l-1}du,
$$
where
$$\eno{
&f(u)=(1+u)^{2l+2a_1-a_2}\prod_{k\ne m_i}\left({1-u^{n_k}\over 1-u}\right)^{
2a_{k-1}-a_k-a_{k-2}}\cdot\cr
&\cdot\prod^{\al}_{i=1}\left({1-u^{y_i}\over 1-u}\right)^{
2a_{m_i-1}-a_{m_i-2}-a_{m_{i+1}}}\left({1-u^{y_{\al +1}}\over 1-u}\right)^{
a_{m_{\al +1}}+a_{m_{\al +1}-1}-a_{m_{\al +1}-2}}.}
$$}

Proof. We shall divide the proof into few steps.

{\bf I step.} Let us put $m_{\al +1}=m$. We define a sequence of formal power
series
$\varphi_1,\ldots ,\varphi_m$ in variables $z_1,\ldots ,z_m,z_0$ by the
following rules:
$$\eno{
&\varphi_m(z_m)=(1-z_m)^{-(a_m+1)}(1-z_0(1-z_m)^{-1})^{-1},\cr
&\varphi_{m-1}(z_{m-1},z_m)=(1-z_{m-1})^{-(a_{m-1}+1)}\varphi_m((1-z_{m-1})^{-1}z_m),\cr
&.................................................................\cr
&\varphi_k(z_k,\ldots ,z_m)=(1-z_k)^{-(a_k+1)}\varphi_{k+1}(
(1-z_k)^{-2b_{k,k+1}}\cdot\cr
&\hskip 2.5cm\cdot z_{k+1},\ldots ,(1-z_k)^{-2b_{k,l}}z_l,\ldots
,(1-z_k)^{-2b_{k,m}}z_m),\cr
&..................................................................\cr
&\varphi_1(z_1,\ldots ,z_m)=(1-z_1)^{-(a_1+1)}\varphi_2((1-z_1)^{-2b_{1,2}}
z_2,\ldots ,(1-z_1)^{-2b_{1,l}}\cdot\cr
&\hskip 2.5cm\cdot z_l,\ldots ,(1-z_1)^{-2b_{1,m}}z_m).}
$$

{\bf Lemma 4.2.} {\it In the power series $\varphi_1(z_1,\ldots ,z_m)$ a
coefficient before $z_o^{\nu_0}z_1^{\nu_1}\ldots z_m^{\nu_m}$ is equal to}
$$\prod_{j=1}^{m-1}\pmatrix{P_j(\nu)+\nu_j\cr
\nu_j}\cdot\pmatrix{a_m+\nu_m+\nu_0\cr
\nu_m}.
$$

Proof.
$$\varphi_m(z_m)=\sum_{\nu_0,\nu_m}z_0^{\nu_0}z_m^{\nu_m}\pmatrix{a_m+\nu_m+\nu_0\cr\nu_m}.$$
Let us assume that
$$\varphi_k(z_k,\ldots ,z_m)=\sum_{\nu_0,\nu_k,\ldots ,\nu_m}A_k(\nu_k,\ldots
,\nu_m;\nu_0)z_0^{\nu_0}z_k^{\nu_k}\ldots z_m^{\nu_m},
$$
then \hskip 0.5cm$\varphi_{k-1}(z_{k-1},\ldots ,z_m)=$
$$\eno{
&=(1-z_{k-1})^{-(a_{k-1}+1)}\varphi_k((1-z_k)^{-2b_{k,k+1}}z_{k+1},\ldots
,(1-z_k)^{-2b_{k,m}}z_m)=\cr
&=\sum_{\nu_0,\nu_k,\ldots ,\nu_m}A_k(\nu_k,\ldots ,\nu_m;\nu_0)(1-z_{k-1})^{
-(p_{k-1}(\nu )+1)}z_0^{\nu_0}z_k^{\nu_k}\ldots z_m^{\nu_m}=\cr
&=\sum_{\nu_0,\nu_{k-1},\ldots ,\nu_m}A_k(\nu_k,\ldots ,\nu_m;\nu_0)\pmatrix{
P_{k-1}(\nu )+\nu_{k-1}\cr\nu_{k-1}}
z_0^{\nu_0}z_{k-1}^{\nu_{k-1}}\ldots z_m^{\nu_m}.}
$$
Consequently,
$$A_{k-1}(\nu_{k-1},\nu_k,\ldots ,\nu_m;\nu_0)=A_k(\nu_k,\ldots ,\nu_m;\nu_0)
\cdot\pmatrix{P_{k-1}(\nu)+\nu_{k-1}\cr\nu_{k-1}}.
$$
\qed

{}From Lemma 4.2 it follows that the sum $Z(\{ a\}|l)$ is equal to the
coefficient before $t^l$ in the power series of $\psi (z,t)$, which has been
obtained
from $\varphi_1(z_1,\ldots ,z_m)$ after substitution
$$\eno{
&z_j=t^{n_j},~~j\ne m-1,\cr
&z_{m-1}=t^{n_{m-1}}z_0^{-1}.}
$$

{\bf II step.} Calculation of the power series for $\psi (z,t)$.

Let us define
$$\eno{
&z_k^{(l)}:=(1-z_l^{(l-1)})^{-2b_{l,k}}\cdot z_k^{(l-1)},~~l\ge 1,
&(4.1)\cr
&z_k^{(0)}=t^{n_k},~~{\rm if}~~k\ne m-1~~{\rm
and}~~z_{m-1}^{(0)}=t^{n_{m-1}}z_0^{-1}.}
$$
Then we have
$$\eno{
&\varphi_1(z_1,\ldots ,z_m)=(1-z_1)^{-(a_1+1)}\varphi_2(z_2^{(1)},z_3^{(1)},
\ldots ,z_m^{(1)})=\cr
&=(1-z_1)^{-(a_1+1)}(1-z_2^{(1)})^{-(a_2+1)}\varphi_3(z_3^{(2)},z_4^{(2)},
\ldots ,z_m^{(2)})=\ldots & (4.2)\cr
&\prod^{m-1}_{j=1}(1-z_j^{(j-1)})^{-(a_j+1)}\cdot
\varphi_{m-1}(z_{m-1}^{(m-2)}, z_m^{(m-2)}).}
$$
In order to compute a formal series $z_k^{(l)}$, we define (see e.g. [K1])
a sequence of polynomials $Q_m(t)$ using the following recurrence relation
$$\eno{
&Q_{m+1}(t)=Q_m(t)-tQ_{m-1}(t),~~m\ge 0,\cr
&Q_0(t)=Q_{-1}(t):=1.}
$$
\medbreak

{\bf Lemma 4.3.} (Formulae for power series $z_k^{(l)}$).

{\it Let us assume that $m_i\le k<m_{i+1}$ and put $m_0:=1$. Then we have
($Q_k:=Q_k(t)$)

1.~~$z_k^{(k-1)}=Q_{k-1}^{-2}Q_{m_i-2}z_k^{(0)}$,

2.~~$1-z_k^{(k-1)}=Q_kQ_{k-1}^{-2}Q_{k-2}$,  if $k\ne m_i$.

3.~~If $k=m_i,~i\ge 1$, then $1-z_k^{(k-1)}=Q_kQ_{k-1}^{-2}Q_{m_{i-1}-2}$.

4.~~After  specialization $t:={\displaystyle{u\over (1+u)^2}}$ one  can find
(remind  that $m_i\le k<m_{i+1}$)
$$Q_k(u)=1-{1-u^{n_k+2y_i}\over (1-u)(1+u)^{n_k+2y_i-1}}.
$$

5.~~If $k\ne m_i+1$ and $m_i\le k<m_{i+1}$, then}
$$z_k^{(k-2)}=Q_{k-3}^2Q_{k-2}^{-4}Q_{m_i-2}^2z_k^{(0)}.
$$

Proof follows by induction from (4.1) and properties of polynomials
$Q_k(t)$ (compare [K1], Lemma 2).
\qed
\medbreak

{\bf Corollary 4.4.}

{\it 1. $z_m^{(m-2)}=Q_{m-3}^2Q_{m-2}^{-2}t^{n_m}$,
{}~~~~$z_{m-1}^{(m-2)}=Q_{m-2}^{-2}Q^2_{m_{\alpha -2}}t^{n_{m-1}}z_0^{-1}$.

2. Let us denote by $\varphi_{m-1}(u,z_0)$ a specialization $t={\displaystyle
{u\over (1-u)^2}}$ of formal series\break
$\varphi_{m-1}(z_{m-1}^{(m-2)},z_m^{(m-1)})$ and
let $\varphi_{m-1}(u)$ be a constant term of series $\varphi_{m-1}(u,z_0)$
w.r.t. variable $z_0$. Then}
$$\varphi_{m-1}(u)=(1-u^{y_{\alpha +1}})^{a_m+a_{m-1}+1}(1-u^{y_{\alpha +1}
-y_{\alpha}})^{-(a_{m-1}+1)}(1-u^{y_{\alpha}})^{-(a_m+1)}.
$$
Remind that $m=m_{\alpha +1}$.
\medbreak

{\bf III step.} Combining (4.2), Lemma 4.3 and Corollary 4.4 after some
simplifications we obtain a proof of Theorem 4.1.

\qed
\medbreak

{\bf Corollary 4.5.} (Combinatorial completeness of Bethe's
states for $XXZ$ model of arbitrary spins).
$$Z=\prod_m(2s_m+1)^{N_m}.\eqno (4.3)
$$

Examples below give an illustration to our result about completeness of Bethe's
states for spin-${1\over 2}$ $XXZ$ - model (Example 1 and 3) and for spin-1
$XXZ$ model (Example 4).
\bigbreak

{\bf Example 1.} We compute firstly the quantities $q_j,~a_j$ (see (3.8))
and after this consider a numerical example. From (3.4) - (3.6) and (3.8) it
follows
$$q_j=(-1)^i~{p_0-n_jp_{i+1}\over y_i}.
$$
Using Theorem 3.2 n.5 we obtain (see (3.9))
$$\eno{&a_j=(-1)^{i-1}n_j\left[{\sum 2s_mN_m-2l\over p_0}\right] +(-1)^{i}
(n_j+q_j)\left({\sum 2s_mN_m-2l\over p_0}\right)+& (4.4)\cr \cr
&+{n_j\over p_0}\sum_{\{m:2s_m\le n_j\}}N_m\left({p_{i+1}\over y_i}(2s_m+1)+
(-1)^iq_{\chi}\right) +\sum_{\{m:2s_m\le n_j\}}N_m\left( 1-{1\over y_i}(2s_m+1)
\right).}
$$

Let us consider the case when all spins are equal to ${1\over 2}$ and let $N$
be the number of spins, then

$i)$ $0\le j<m_1~(=\nu_0)$. Then $r(j)=i=0$ and $n_j=j,~~q_j=p_0-j,$
$$a_j=-n_j\left[{N-2l\over p_0}\right] +N-2l+\delta_{n_j,1}{N\over p_0}(2-p_0+
q_{\chi}).
$$

$ii)$ $m_1\le j<m_2~(=\nu_0+\nu_1)$. Then $r(j)=1$ and $n_j=1+(j-m_1)
\nu_0$,

{}~~~~~$q_j=(p_0-\nu_0)(j-m_1)-1$,
$$a_j=n_j\left[{N-2l\over p_0}\right] -{N-2l\over \nu_0}~(n_j-1)-\delta_{n_j,1}
{N\over p_0}(2-p_0+q_{\chi}).
$$
For example, $a_{m_1}=\left[{\displaystyle{N-2l\over p_0}}\right]-
\displaystyle{N\over p_0}(2-p_0+q_{\chi})$.
\vskip 0.3cm

$iii)$ $m_2\le j<m_3~(=\nu_0+\nu_1+\nu_2)$. Then $r(j)=2$ and $n_j=\nu_0+
(j-m_2)(1+\nu_0\nu_1)$,

{}~~~~~~$q_j=p_0-\nu_0-(j-m_2)(1-\nu_1(p_0-\nu_0))$
$$a_j=-n_j\left[{N-2l\over p_0}\right] +{N-2l\over\nu_0+\displaystyle{1
\over\nu_1}}~(n_j+\displaystyle{1\over \nu_1}).
$$
Consequently,
$$a_{m_2}=-\nu_0\left[{N-2l\over p_0}\right ]+(N-2l).
$$

Now let us assume $p_0=3+{\displaystyle{1\over 3}},~~N=5$. It is clear that in
our
case $\chi =2$ (see (3.7)) and $q_{\chi}=p_0-2$.
Below we give all solutions $\ld =\{\ld_1,\ld_2,\ldots\}$ to the equation
(3.11)
when $0\le l\le 2$ and compute the corresponding vacancy numbers
$P_j=P_j(\ld )$
(see (3.9)) and number of states $Z=Z(N,~{\displaystyle{1\over 2}}~|~\{\ld_k\}
)$
(see (3.10) and (3.12)):
$$\matrix{l=0&&\{ 0\}&&P_j=0&&Z=1\cr &&&&&&&&Z(5,{1\over 2}~|~0)=1\cr
l=1&&\{ 1,0,0\}&&P_1=3&&Z=4\cr
&&\{ 0,0,1\}&&P_3=0&&Z=1\cr &&&&&&&&Z(5,{1\over 2}~|~1)=5\cr
l=2&&\{ 0,1,0\}&&P_2=1&&Z=2\cr
&&\{ 2,0,0\}&&P_1=1&&Z=3\cr &&\{ 0,0,2\}&&P_3=0&&Z=1\cr &&\{ 1,0,1\}
&&\cases{P_1=3\cr P_3=0}&&Z=4\cr&&&&&&&&Z(5,{1\over 2}~|~2)=10}
$$
Consequently,
$$Z(N=5,{1\over 2})=2(1+5+10)=32=2^5.$$

\def\wdt{\widetilde}
Note that our formula (3.10) for the number of Bethe's states with fixed spin
$l$, namely $Z(N,s~|~l)$, works for $l\ge \sum s_mN_m$ as well as for small
$l\le\sum s_mN_m$.

In Appendix we consider two additional examples, one when all spins are equal
to ${1\over 2}$, another one when all spins are equal to $1$. The last example
seems to be interesting because nonadmissible configuration appears.
\medbreak

{\bf Remark 1.} It is easy to see that for fixed $l$ and sufficiently big
$N=\sum 2s_mN_m$ all vacancy numbers $P_j(\ld )$ are non negative. It is not
the case for particular $N$ and we must consider really the configurations with
$$P_j(\ld )+\ld_j <0~~{\rm for~some}~~j\eqno (4.5)
$$
in order to have a correct answer for $Z^{XXZ}(N,s~|~l)$. See
Appendix, Example 4, $l=4$, $(\clubsuit )$.
Let us remind that for $XXX$ model the nonadmissible
configurations(i.e. those satisfying (4.5)) give a zero contribution to the
sum $Z^{XXX}(N,s~|~l)$,
[K2].
\medbreak

{\bf Remark 2.} One can rewrite the expressions (3.9) for vacancy numbers in
the following form if $m_i\le j<m_{i+1}$
$$\eno{P_j(\ld )&=(-1)^{i-1}\left(\sum_m2\Phi_{j,2s_m}\cdot N_m-n_j\left\{
{\sum 2s_mN_m-2l\over p_0}\right\}\right)-\cr
&-\sum_k 2(-1)^{r(k)}{\widetilde\theta_{jk}}\ld_k
-\delta_{j,m_{\al +1}-1}\ld_{m_{\al +1}}+
\delta_{j,m_{\al +1}}\ld_{m_{\al +1}-1},}
$$
where ${\wdt\theta_{jk}}=(-1)^{r(j)+r(k)}\displaystyle{n_jq_k\over p_0}$, if
$j\le k$ and ${\wdt\theta_{jk}}={\wdt\theta_{kj}}$.

Let us introduce the symmetric matrix $\Theta =({\wdt\theta_{ij}})_{1\le i,j
\le m_{\al +1}}$. We can find the inverse matrix $\Theta^{-1}:=(c_{ij})$
and compute its determinant.
\medbreak

{\bf Theorem 4.6.} {\it Matrix elements $c_{ij}$ of the inverse matrix
$\Theta^{-1}$ are given by the following rules}

$i)~~~~c_{ij}=c_{ji}$ \  and \ $c_{ij}=0$, \ if \ $|i-j|\ge 2$.

$ii)$~~~$c_{j-1,j}=(-1)^{i-1}$, if $m_i\le j<m_{i+1}$.
\vskip 0.3cm

$iii)$~~$c_{jj}=\cases{2(-1)^i,&if $m_i\le j<m_{i+1}-1,~~i\le\al$,\cr
(-1)^i,& if $j=m_{i+1}-1,~~i\le\al$,\cr
(-1)^{\al +1},& if $j=m_{\al +1}$}$
\vskip 0.5cm

\medbreak
{\bf Theorem 4.7.} {\it We have}
$${\rm det}|\Theta^{-1}|=y_{\al +1}
$$

The proofs of Theorems 4.6 and 4.7 follow from [KR2], Appendix A, and
relations
$$y_ip_i+y_{i-1}p_{i+1}=p_0,~~0\le i\le\al +1.
$$
\qed

{\bf Example 2.} For $p_0=4+\displaystyle{1\over 5}$ using Theorem 4.6
one can find
$$\Theta^{-1}=\pmatrix{2&-1\cr -1&2&-1\cr &-1&1&1\cr &&1&-2&1\cr &&&1&-2&1\cr
&&&&1&-2&1\cr &&&&&1&-2&1\cr &&&&&&1&-1&-1\cr &&&&&&&-1&1}
$$
\vskip 0.3cm

{\bf \S 5. Conclusion.}
\bigbreak

In our paper we have proved a very general combinatorial identity (Theorem
4.1).
As a particular case we proved a combinatorial completeness of Bethe's states
for generalized $XXZ$ model (Corollary 4.5). One can construct a natural
$q$-analog for the number of Bethe's states with fixed spin $l$ (see (3.10)).
Namely, let us consider a vector
$$
{\wdt\ld}=({\wdt\ld_1},\ldots ,{\wdt\ld_{m_{\al +1}}}),
$$
where ${\wdt\ld_j}=(-1)^{r(j)}\ld_j$
and a matrix $E=(e_{jk})_{1\le j,k\le
m_{\al +1}}$, where
$$e_{jk}=(-1)^{r(k)}(\delta_{j,k}-\delta_{j,m_{\al +1}-1}\cdot\delta_{k,m_{\al
+1}}+\delta_{j,m_{\al +1}}\cdot\delta_{k,m_{\al +1}-1}).
$$
Then it is easy to check that
$$P_j(\ld )+\ld_j=((E-2\Theta ){\wdt\ld^t}+b^t)_j,
$$
where $b=(b_1,\ldots ,b_{m_{\al +1}})$ and
$$b_j=(-1)^{r(j)}\left(n_j\left\{{\sum 2s_mN_m-2l\over p_0}\right\}
-\sum_m2\Phi_{j,2s_m}\cdot N_m\right).
$$
We consider the following $q$-analog of (3.10)
$$\sum_{\ld}q^{{1\over 2}{\wdt\ld}B{\wdt\ld^t}}\prod_j\left[\matrix{
((E-B){\wdt\ld}^t+b^t)_j\cr \ld_j}\right]_q,\eqno (5.1)
$$
where summation is taken over all configurations $\ld =\{\ld_k\}$ such that
$$\sum_{k=1}^{m_{\al +1}}n_k\ld_k=l,~~\ld_k\ge 0,~~{\rm and}~~B=2\Theta .
$$
The thermodynamical limit of (5.1) (i.e. $N_m\to\infty$) comes to
$$\sum_{\ld}{q^{{1\over 2}{\wdt\ld}B{\wdt\ld^t}}\over\displaystyle{\prod_j
(q)_{\ld_j}}},\eqno (5.2)
$$
summation in (5.2) is the same as in (5.1) and $(q)_n:=(1-q)\cdots (1-q^n)$.
Here $B=C_1\otimes\Theta$ and $C_1=(2)$ is the Cartan matrix of type
$A_1$.

It is an interesting problem to find a representation theory meaning of (5.2),
when $B=C_k\otimes\Theta$ and $C_k$ is the Cartan matrix of type $A_k$.

Another interesting question is that about a degeneration of Bethe's states
for $XXZ$ model into those for $XXX$ one. More exactly, we had proved (see
(4.3)) that
$$\prod_m(2s_m+1)^{N_m}=\sum^N_{l=0}Z^{XXZ}(N,s~|~l),\eqno (5.3)
$$
where $N=\sum_m2s_mN_m$ and $Z^{XXZ}(N,s~|~l)$ is given by (3.10).

On the other hand it follows from a combinatorial completeness of Bethe's
states
for $XXX$ model (see [K1]) that
$$\prod_m(2s_m+1)^{N_m}=\sum_{l\ge 0}^{{1\over 2}N}(N-2l+1)Z^{XXX}(N,s~|~l),
\eqno (5.4)
$$
where $Z^{XXX}(N,s~|~l)$ is the multiplicity of $({N\over 2}-l)$ - spin
irreducible representation $V_{{N\over 2}-l}$ of $sl(2)$ in the tensor product
$$V_{s_1}^{\otimes N_1}\otimes\cdots\otimes V_{s_m}^{\otimes N_m}.
$$
It is an interesting question to find a combinatorial proof that
$${\rm RHS} (5.3)={\rm RHS} (5.4).
$$
Another interesting task is to compare our results with those obtained in [KM].
We assume to consider this questions and also to study in more details the case
{}~$p_0=\nu_0$~ is an integer and all spins are equal to \
$\displaystyle{\nu_0-2
\over 2}$ \ in the separate publications.
\bigbreak

{\bf Acknowledgements.} We are pleased to thank for hospitality our colleagues
from Tokyo University, where this work was completed.
\vfill\eject

{\bf Appendix.}
\bigbreak

{\bf Example 3.} Using the same notations as in Example 1, we consider the
case ${s={1\over 2}}$,\break $p_0=\displaystyle{3+{1\over 3}},~~N=8$ and find
the
vacancy numbers $P_j(\ld )$ and numbers of states\break
$Z=Z(N,{1\over 2}~|~\{\ld_k\})$:
$$\matrix{l=0&&\{ 0\}&&P_j=0&&Z=1\cr &&&&&&&&Z(8,{1\over 2}~|~0)=1\cr
l=1&&\{ 1,0,0\}&&P_1=5&&Z=6\cr&&\{ 0,0,1\}&&P_3=1&&Z=2\cr
&&&&&&&&Z(8,{1\over 2}~|~1)=8\cr
l=2&&\{ 2,0,0\}&&P_1=3&&Z=10\cr &&\{ 0,0,2\}&&P_3=1&&Z=3\cr
&&\{ 0,1,0\}&&P_2=2&&Z=3\cr &&\{ 1,0,1\}&&\cases{P_1=5\cr P_3=1}&&Z=12\cr
&&&&&&&&Z(8,{1\over 2}~|~2)=28\cr
l=3&&\{ 3,0,0\}&&P_1=2&&Z=10\cr &&\{ 0,0,3\}&&P_3=0&&Z=1\cr
&&\{ 0,0,0,0,0,1\}&&P_6=2&&Z=3\cr &&&&&&&&Z(8,{1\over 2}~|~3)=56\cr
&&\{ 1,1,0\}&&\cases{P_1=4\cr P_2=2}&&Z=15\cr
&&\{ 0,1,1\}&&\cases{P_2=4\cr P_3=0}&&Z=5\cr
&&\{ 2,0,1\}&&\cases{P_1=4\cr P_3=0}&&Z=15\cr
&&\{ 1,0,2\}&&\cases{P_1=6\cr P_3=0}&&Z=7\cr
l=4&&\{ 4,0,0\}&&P_1=0&&Z=1\cr &&\{ 0,0,4\}&&P_3=0&&Z=1\cr
&&\{ 0,2,0\}&&P_2=0&&Z=1\cr &&\{ 0,0,0,1\}&&P_4=0&&Z=1\cr
&&\{ 2,1,0\}&&\cases{P_1=2\cr P_2=0}&&Z=6\cr
&&\{ 0,1,2\}&&\cases{P_2=4\cr P_3=0}&&Z=5\cr
&&\{ 3,0,1\}&&\cases{P_1=2\cr P_3=0}&&Z=10\cr
&&\{ 1,0,3\}&&\cases{P_1=6\cr P_3=0}&&Z=7\cr
&&\{ 2,0,2\}&&\cases{P_1=4\cr P_3=0}&&Z=15\cr
}$$
$$\matrix{
{}~~~&&\{ 1,0,0,0,0,1\}&&\cases{P_1=4\cr P_6=0}&&Z=5\cr
&&\{ 0,0,1,0,0,1\}&&\cases{P_3=2\cr P_6=0}&&Z=3\cr
&&\{1,1,1,0,0,0\}&&\cases{P_1=4\cr P_2=2\cr P_3=0}&&Z=15\cr
&&&&&&&&Z(8,{1\over 2}~|~4)=70}
$$
Consequently,
$$Z(N=8,{1\over 2}~|~l)=\pmatrix{8\cr l},~~~0\le l\le 4,
$$
and
$$Z(N=8,{1\over 2})=2(1+8+28+56)+70=256=2^8.
$$

{\bf Example 4.} Let us consider the case when all spins are equal to $1$ and
let $N$ be the number of spins. We compute firstly the quantities $a_j$
(see (3.8)) and after this consider a numerical example.

$i)$ $0\le j<m_1~(=\nu_0)$. Then $r(j)=i=0$ and $n_j=j,~~q_j=p_0-j,$
$$a_j=\cases{-j\left[\displaystyle{2N-2l\over p_0}\right] +2N-2l, &if $j>2$\cr
\cr -j\left[\displaystyle{2N-2l\over p_0}\right] +\displaystyle{jN\over p_0}
(3+q_{\chi})-2l, &if $j\le 2$}
$$

$ii)$ $m_1\le j<m_2~(=\nu_0+\nu_1)$. Then $r(j)=1$ and $n_j=1+(j-m_1)
\nu_0$,

{}~~~~~$q_j=(p_0-\nu_0)(j-m_1)-1$,
$$a_j=n_j\left[{2N-2l\over p_0}\right] -{2N-2l\over
\nu_0}(n_j-1)-\delta_{n_j,1}
{N\over p_0}(3-p_0+q_{\chi}).
$$

$iii)$ $m_2\le j<m_3~(=\nu_0+\nu_1+\nu_2)$. Then $r(j)=2$ and
$n_j=\nu_0+(j-m_2)(1+\nu_0\nu_1)$,

{}~~~~~~$q_j=p_0-\nu_0-(j-m_2)(1-\nu_1(p_0-\nu_0))$,
$$a_{m_2}=-\nu_0\left[{2N-2l\over p_0}\right] +2N-2l.
$$

Now let us assume $p_0=3+{\displaystyle{1\over 3}},~~N=5$. It is clear that
$\chi =6$
and $q_{\chi}={1\over 3}$.
Below we give all solutions $\ld =\{\ld_1,\ld_2,\ldots\}$ to the equation
(3.11)
when $0\le l\le 5$ and compute the corresponding vacancy numbers
$P_j=P_j(\ld )$
(see (3.9)) and number of states\break
$Z=Z(N,1~|~\{\ld_k\} )$ (see (3.10) and (3.12)):

$$\matrix{l=0&&\{ 0\}&&P_j=0&&Z=1\cr
&&&&&&&&Z(5,1~|~0)=1\cr
l=1&&\{ 1,0,0\}&&P_1=1&&Z=2\cr
&&\{ 0,0,1\}&&P_3=2&&Z=3\cr
&&&&&&&&Z(5,1~|~1)=5\cr
l=2&&\{ 2,0,0\}&&P_1=0&&Z=1\cr
&&\{ 0,0,2\}&&P_3=1&&Z=3\cr
&&\{ 0,1,0\}&&P_2=4&&Z=5\cr &&\{ 1,0,1\}&&\cases{P_1=2\cr P_3=1}&&Z=6\cr
&&&&&&&&Z(5,1~|~2)=15\cr
l=3&&\{ 3,0,0\}&&P_1=-2&&Z=0\cr
&&\{ 0,0,3\}&&P_3=1&&Z=4\cr
&&\{ 0,0,0,0,0,1\}&&P_6=1&&Z=2\cr
&&\{ 1,1,0\}&&\cases{P_1=0\cr P_2=2}&&Z=3\cr
&&\{ 0,1,1\}&&\cases{P_2=4\cr P_3=1}&&Z=10\cr
&&\{ 2,0,1\}&&\cases{P_1=0\cr P_3=1}&&Z=2\cr
&&\{ 1,0,2\}&&\cases{P_1=2\cr P_3=1}&&Z=9\cr
&&&&&&&&Z(5,1~|~3)=30\cr
l=4&&\{ 4,0,0\}&&P_1=-3&&Z=0\cr
&&\{ 0,0,4\}&&P_3=0&&Z=1\cr
&&\{ 0,2,0\}&&P_2=2&&Z=6\cr
&&\{ 0,0,0,1\}&&P_4=-2&&Z=-1&&(\clubsuit )~~~~~~~~~~~\cr
&&\{ 2,1,0\}&&\cases{P_1=-1\cr P_2=2}&&Z=0\cr
&&\{ 0,1,2\}&&\cases{P_2=6\cr P_3=0}&&Z=7\cr
&&\{ 3,0,1\}&&\cases{P_1=-1\cr P_3=0}&&Z=0\cr
&&\{ 1,0,3\}&&\cases{P_1=3\cr P_3=0}&&Z=4\cr
&&\{ 2,0,2\}&&\cases{P_1=1\cr P_3=0}&&Z=3\cr
&&\{ 1,0,0,0,0,1\}&&\cases{P_1=1\cr P_6=2}&&Z=6\cr
&&\{ 0,0,1,0,0,1\}&&\cases{P_3=2\cr P_6=2}&&Z=9\cr
&&\{ 1,1,1,0,0,0\}&&\cases{P_1=1\cr P_2=4\cr P_3=0}&&Z=10\cr
&&&&&&&&Z(5,1~|~4)=45\cr
}$$
$$\matrix{
l=5&&\{ 5,0,0\}&&P_1=-5&&Z=0\cr
&&\{0,0,5\}&&P_3=0&&Z=1\cr
&&\{ 4,0,1\}&&\cases{P_1=-3\cr P_3=0}&&Z=0\cr
&&\{ 1,0,4\}&&\cases{P_1=3\cr P_3=0}&&Z=4\cr
&&\{ 3,0,2\}&&\cases{P_1=-1\cr P_3=0}&&Z=0\cr
&&\{ 2,0,3\}&&\cases{P_1=1\cr P_3=0}&&Z=3\cr
&&\{ 3,1,0\}&&\cases{P_1=-3\cr P_2=0}&&Z=0\cr
&&\{ 0,1,3\}&&\cases{P_2=6\cr P_3=0}&&Z=7\cr
&&\{ 1,2,0\}&&\cases{P_1=-1\cr P_2=0}&&Z=0\cr
&&\{ 0,2,1\}&&\cases{P_2=2\cr P_3=0}&&Z=6\cr
&&\{ 1,0,0,1\}&&\cases{P_1=1\cr P_4=0}&&Z=2\cr
&&\{ 0,0,1,1\}&&\cases{P_3=2\cr P_4=0}&&Z=3\cr
&&\{ 0,1,0,0,0,1\}&&\cases{P_2=2\cr P_6=0}&&Z=3\cr
&&\{ 2,0,0,0,0,1\}&&\cases{P_1=-1\cr P_6=0}&&Z=0\cr
&&\{ 0,0,2,0,0,1\}&&\cases{P_3=2\cr P_6=0}&&Z=6\cr
&&\{ 1,0,1,0,0,1\}&&\cases{P_1=1\cr P_3=2\cr P_6=0}&&Z=6\cr
&&\{ 2,1,1\}&&\cases{P_1=-1\cr P_2=2\cr P_3=0}&&Z=0\cr
&&\{ 1,1,2\}&&\cases{P_1=1\cr P_2=4\cr P_3=0}&&Z=10\cr
&&&&&&&&Z(5,1~|~5)=51}
$$
$$Z(N=5,1)=2(1+5+15+30+45)+51=243=3^5.
$$
\vfill\eject

{\bf References.}
\medbreak

\item{[TS]} Takahashi M., Suzuki M. One-dimensional anisotropic Heisenberg
model
at finite temperatures, Progr. of Theor. Phys., 1972, v. 48, N. 6B,
p. 2187-2209.

\item{[K1]} Kirillov A.N. Combinatorial identities and completeness of states
for
the generalized Heisenberg magnet, Zap. Nauch. Sem. LOMI, 1984, v. 131,
p. 88-105.

\item{[K2]} Kirillov A.N. On the Kostka-Green-Foulkes polynomials and
Clebsch-Gordon numbers, Journ. Geom. and Phys., 1988, v. 5, N. 3,
p. 365-389.

\item{[KR1]} Kirillov A.N., Reshetikhin N.Yu. Properties of kernels of
integrable
equations for $XXZ$ model of arbitrary spin, Zap. Nauch. Sem. LOMI,
1985, v. 146, p. 47-91.

\item{[KR2]} Kirillov A.N., Reshetikhin N.Yu. Exact solution of the integrable
$XXZ$ Heisenberg model with arbitrary spin, J. Phys. A.: Math. Gen.,
1987, v. 20, p. 1565-1597.

\item{[FT]} Faddeev L.D., Takhtadjan L.A. Spectrum and scattering of
excitations in one dimensional isotropic Heisenberg model, Zap. Nauch. Sem.
LOMI, 1981, v. 109, p. 134.

\item{[EKK]} Essler F., Korepin V.E., Schoutens K. Fine structure of the Bethe
ansatz for the\break spin-${1\over 2}$ Heisenberg $XXX$ model, J. Phys. A:
Math. Gen.,
1992, v. 25, p. 4115-4126.

\item{[KM]} Kedem R., McCoy B. Construction of modular branching functions
from Bethe's equations in the 3-state Potts chain, J. Stat. Phys., 1993,
v. 74, p. 865.

\end